\documentclass[aps,prb,showpacs,twocolumn]{revtex4}

\usepackage[T1]{fontenc}
\usepackage{amssymb}
\usepackage{amsbsy}
\usepackage{amsmath}
\usepackage{graphicx}
\usepackage{hyperref} 

\begin{document}

\def\beq{\begin{equation}}
\def\eeq{\end{equation}}
\def\bleq{\begin{eqnarray}}
\def\eleq{\end{eqnarray}} 
\newcommand{\Tr}{{\rm Tr}} 
\newcommand{\tr}{{\rm tr}} 
\newcommand{\mean}[1]{\langle #1 \rangle}
\newcommand{\const}{{\rm const}} 
\newcommand{\ie}{i.e. }
\newcommand{\eg}{e.g. }
\newcommand{\cc}{{\rm c.c.}} 
\def\eps{\epsilon}
\def\half{\frac{1}{2}}
\def\q{{\bf q}}
\def\r{{\bf r}}
\def\nablabf{\boldsymbol{\nabla}}
\def\dt{\partial_t}

\title{Non-perturbative renormalization-group approach to lattice models}
  
\author{N. Dupuis}
\affiliation{Laboratoire de Physique Th\'eorique de la Mati\`ere Condens\'ee, 
CNRS - UMR 7600, \\ Universit\'e Pierre et Marie Curie, 4 Place Jussieu, 
75252 Paris Cedex 05,  France }

\affiliation{Laboratoire de Physique des Solides, CNRS - UMR 8502,
  Universit\'e Paris-Sud, 91405 Orsay, France}

\author{K. Sengupta}
\affiliation{TCMP division, Saha Institute of Nuclear
Physics, 1/AF Bidhannagar, Kolkata-700064, India}

\affiliation{Theoretical Physics Department, Indian Association for the Cultivation of Science, Jadavpur, Kolkata 700 032, India} 

\date{October 13, 2008}

\begin{abstract}
The non-perturbative renormalization-group approach is extended to lattice models, considering as an example a $\phi^4$ theory defined on a $d$-dimensional hypercubic lattice. Within a simple approximation for the effective action, we solve the flow equations and obtain the renormalized dispersion $\eps(\q)$ over the whole Brillouin zone of the reciprocal lattice. 
In the long-distance limit, where the lattice does not matter any more, we reproduce the usual flow equations of the continuum model. We show how the numerical solution of the flow equations can be simplified by expanding the dispersion in a finite number of circular harmonics. 
\end{abstract}


\maketitle

\section{Introduction}

The non-perturbative renormalization group (NPRG) has proven to be a powerful tool in the study of systems with a large number of interacting degrees of freedom.\cite{Wilson74,Polchinski84,Wetterich93} It has been successfully applied in many areas of physics, condensed matter, nuclear and particle physics, etc. (For a review and a pedagogical introduction, see Refs.~\onlinecite{Berges00} and \onlinecite{Delamotte07}.) 

The NPRG is based on an exact flow equation satisfied by the effective action $\Gamma[M]$ (\ie the generating functional of one-particle irreducible vertices). In the most common approximation, one expands the effective action in powers of derivatives of the field $M$. While this approach is often the easiest way to solve the NPRG equations, it yields only the small momentum behavior of the vertices.\cite{note1} As shown recently, it is possible to solve the NPRG equations beyond the derivative expansion and compute the whole momentum dependence of the vertices.\cite{Blaizot06,Blaizot06a,Blaizot06b,Guerra07,Blaizot07,Benitez08,Hasselmann07,Sinner08} 

A proper description of the momentum dependence of the vertices opens up the possibility to study lattice models. While the derivative expansion is always appropriate to study the long-distance physics and predicts universal quantities such as the critical exponents, it often fails to predict non-universal quantities (\eg the critical temperature of a phase transition) which depend on the short-distance physics. Moreover, in some cases the lattice is the very reason for the occurrence of a phase transition.\cite{note2} (For approximate treatments of lattice models within the derivative expansion, see Refs.~\onlinecite{Baier04,Baier05,Krahl07,Krahl08}.) 

In this paper, we show how the usual NPRG approach should be modified in the case of lattice models. As an example, we consider a $\phi^4$ theory defined on a $d$-dimensional hypercubic lattice. We solve the flow equations within a simple approximation for the effective action. From the 2-leg vertex, we deduce the renormalized ``dispersion'' (\ie the kinetic energy in particle language) over the whole Brillouin zone. In order to simplify the numerical solution of the flow equations, we introduce two approximations: i) the LPA' (where LPA stands for Local Potential Approximation) -- a natural generalization to the lattice case of the LPA' used in continuum models\cite{note3} --, which neglects the renormalization of the dispersion except at very small momenta; ii) the H-LPA' where the LPA' is supplemented by a circular harmonic expansion of the 2-leg vertex, thus allowing one to take into account the renormalization of the dispersion in a (numerically) efficient way. By comparing with the exact solution of the flow equations, these two approximations are found to be remarkably accurate.

\section{NPRG for lattice models}

\subsection{General method}
\label{sec_method}

We consider a $\phi^4$ theory defined on $d$-dimensional hypercubic lattice,
\beq
S[\phi] = \sum_\r \left\lbrace \half \phi_\r [\eps_0(-i\nablabf) + v] \phi_\r + \frac{u}{4!} \phi^4_\r \right\rbrace ,
\label{action}
\eeq
where $\lbrace\r\rbrace$ denotes the sites of the lattice. For simplicity, we consider a one-component real field $\phi_\r$; the extension of our approach to a $N$-component field model with $O(N)$ symmetry is straightforward. The bare dispersion $\eps_0(\q)$ is chosen such that $\eps_0(\q=0)=0$ and $\lim_{\q\to 0}\eps_0(\q)=\eps_0 \q^2$. (This is always possible by a redefinition of the parameter $v$ in (\ref{action}).) For a system with nearest-neighbor interactions only (the case we shall consider for the numerical solution of the flow equations), 
\beq
\eps_0(\q) = -2\eps_0 \sum_{\nu=1}^d [\cos(q_\nu)-1] ,
\label{eps_0_def}
\eeq
where the maximum of $\eps_0(\q)$ is given by $4d\eps_0$. The lattice spacing is taken as the unit length. 

To implement the RG procedure, we add to the action the regulator term
\beq
\Delta S_R[\phi] = \half \sum_\r \phi_\r R(-i\nablabf) \phi_\r = \half \sum_\q \phi_{-\q} R(\q) \phi_\q ,
\label{SR}
\eeq
where the function $R(\q)\equiv R_k(\q)$ depends on the energy $\eps_k$. $\phi_\q=N^{-1/2} \sum_\r e^{-i\q\cdot\r}\phi_\r$ is the Fourier transformed field with $N$ the number of lattice sites. The sum over $\q$ in (\ref{SR}) is restricted to the first Brillouin zone $]-\pi,\pi]^d$ of the reciprocal lattice. In the thermodynamic limit ($N\to\infty$),
\beq
\frac{1}{N} \sum_\q \to \int_{-\pi}^\pi \frac{dq_1}{2\pi} \cdots \int_{-\pi}^\pi \frac{dq_d}{2\pi} \equiv \int_\q .
\eeq 

The NPRG approach is based on the effective action $\Gamma[M]$ defined as the Legendre transform of the free energy $-\ln Z[h]$ from which $\Delta S_R[M]$ is subtracted. Here $h$ denotes an external field that couples linearly to the $\phi$ field and $M_\r=\mean{\phi_\r}_h$ is the expectation value of $\phi_\r$. $\Gamma[M]$ satisfies the exact flow equation\cite{Wetterich93} 
\beq
\partial_k \Gamma[M] = \half \Tr \Bigl\lbrace \partial_k R \bigl(\Gamma^{(2)}[M]+R \bigr)^{-1} \Bigr\rbrace  
\label{eq_exact}
\eeq
as the energy scale $\eps_k$ is varied. $\Gamma^{(2)}[M]$ is the second-derivative of $\Gamma[M]$ with respect to $M$. To keep the notation simple, we do not explicitly indicate the $k$ dependence of $\Gamma$, $\Gamma^{(2)}$ and $R$. 

An important difference with continuum models comes from the regulator function $R$. The latter is chosen to be of the form
\beq
R(\q) = Z \eps_0(\q) r(\tilde\eps_0(\q)), \qquad \tilde \eps_0(\q) = \frac{\eps_0(\q)}{\eps_k} , 
\label{Rdef}
\eeq
where the $k$-dependent variable $Z$ will be specified below. A typical choice for the function $r$ is $r(x) = (e^{x}-1)^{-1}$. The regulator function $R$ gives a mass of order $\eps_k$ to low-energy fluctuation modes ($\eps_0(\q)\lesssim \eps_k$) but leaves the high-energy modes ($\eps_0(\q)\gtrsim \eps_k$) essentially unaffected. When $\eps_k\to \infty$ -- or, in practice, $\eps_k$ larger than any typical energy scale -- all fluctuations are frozen and mean-field theory becomes exact: $\Gamma[M]=S[M]$. As long as $\eps_k\gg \eps_0$ (\ie $\eps_k \gg \eps_0(\q)$), fluctuations are local (on-site). They become non-local when $\eps_k\sim\eps_0$. Although $R(\q)\equiv R_k(\q)$ is a function of $\eps_k$, it is convenient to write $\eps_k=\eps_0k^2$ in terms of a momentum scale $k$ and consider all quantities of interest to be function of $k$ rather than $\eps_k$. 
The regime where fluctuations are local in space then corresponds to length scales $k^{-1}\ll 1$, \ie  much smaller than the lattice spacing, whereas the condition $k^{-1}\gg 1$ implies that fluctuations can propagate through the lattice. Since the function $r(x)$ typically vanishes exponentially for $x\gg 1$,
\beq
\lim_{k\to 0} R(\q) = Z \eps_0 \q^2 r \left(\frac{\q^2}{k^2}\right) , 
\label{Rlimit}
\eeq
and we reproduce the regulator function that is used in continuum models ($\eps_0(\q)=\eps_0\q^2$).\cite{Berges00,Delamotte07} Eq.~(\ref{Rlimit}) expresses the fact that when $\eps_k\ll \eps_0$ (\ie $k\ll 1$) the lattice does not matter any more and only the small-$\q$ limit $\eps_0 \q^2$ of the dispersion plays a role. To distinguish between these two regimes, characterized by the presence or absence of strong lattice effects, it is convenient to introduce the crossover momentum scale $k_x \sim 1$ ($\eps_{k_x} \sim \eps_0$). 

In the following we write $k\equiv k(t)=\Lambda e^{t}$ ($t<0$) where $\Lambda$ is such that when $\eps_k=\eps_0\Lambda^2$, all fluctuations are effectively frozen and the mean-field theory a good approximation. In practice, one should verify that the solution of the flow equations remains essentially unchanged when $\Lambda$ is increased above the chosen value. 

\subsection{Flow equations} 

In this section, we derive the flow equations in the case where the effective action is approximated by the simple form
\beq
\Gamma[M] = \sum_\r \left\lbrace \half M_\r \eps(-i\nablabf) M_\r + U(\rho_\r) \right\rbrace ,
\label{eff_action}
\eeq
where $\eps(\q)$ denotes the ($k$-dependent) dispersion. For symmetry reasons, the potential $U$ can only be a function of $\rho_\r=M_\r^2/2$. In the ordered phase -- the only one that will be of interest to us--, we approximate it by
\beq
U(\rho_\r) = \frac{\lambda}{2} (\rho_\r-\rho_0)^2 ,
\label{Udef}
\eeq
where $\rho_0$ is the $k$-dependent minimum of the potential. The wave-function renormalization factor $Z$ is defined by
\beq
Z = \lim_{\q\to 0} \frac{\eps(\q)}{\eps_0(\q)} = \lim_{\q\to 0} \frac{\eps(\q)}{\eps_0\q^2} . 
\label{Zdef}
\eeq
The dependence of $R$ on $Z$ [Eq.~(\ref{Rdef})] is a necessary condition for the effective action $\Gamma[M]$ to reach a fixed point when the system becomes critical ($\rho_0(k\to 0)=0^+$). We are therefore left with three unknown parameters ($\rho_0$, $\lambda$ and $Z$) and a function $\eps(\q)$ to be determined as a function of $k(t)$. The initial values at $t=0$ are $\rho_0=-3v/u$ (for $v\leq 0$), $\lambda=u/3$ and $\eps(\q)=\eps_0(\q)$, where $u$ and $v$ are introduced in (\ref{action}). 

Given our choice of the regulator function [Eqs.~(\ref{Rdef},\ref{Rlimit})], we expect the flow equations to reproduce those of the continuum model when $k\ll 1$. This suggests to introduce the dimensionless variables
\bleq
\tilde M_\r = (Z \eps_k)^{1/2} k^{-d/2} M_\r , &\quad&
\tilde \rho_\r = Z \eps_k k^{-d} \rho_\r , \nonumber \\ 
\tilde \lambda = Z^{-2} \eps_k^{-2} k^d \lambda , &\quad&
\tilde \eps(\q) = (Z\eps_k)^{-1} \eps(\q) . \nonumber \\ && 
\eleq
The effective action then takes the form
\bleq
\Gamma[\tilde M] &=& k^d \sum_\r \biggl[ \half \tilde M_\r \tilde\eps(-i\nablabf) \tilde M_\r + \frac{\tilde\lambda}{2} \left(\tilde\rho_\r-\tilde\rho_0\right)^2 \biggr] \nonumber \\
&{\simeq \atop {(k\to 0)}}& \int d^d\tilde r \biggl[ \half \left(\nablabf_{\tilde \r} \tilde M(\tilde \r) \right)^2 + \frac{\tilde\lambda}{2} \left(\tilde\rho(\tilde \r)-\tilde\rho_0\right)^2 \biggr] , \nonumber \\ &&
\eleq 
where the last result is obtained in the continuum limit using $\lim_{k\to 0}\tilde\eps(\q)\simeq \q^2/k^2$ and introducing the dimensionless continuous variable $\tilde \r=k\r$. 

The flow equations are deduced from (\ref{eq_exact}) and (\ref{eff_action}). In dimensionless form, one finds (Appendix \ref{app_floweq}) 
\bleq
\dt \tilde\rho_0 &=& -(d-2+\eta)\tilde\rho_0 +3L^d_1(k,2\tilde\lambda\tilde\rho_0,\eta;\tilde\eps) , \nonumber \\ 
\dt \tilde\lambda &=& (d-4+2\eta)\tilde\lambda + 9 \tilde\lambda^2 L^d_2(k,2\tilde\lambda\tilde\rho_0,\eta;\tilde\eps) , \nonumber \\ 
\eta &=& \frac{36 \tilde\lambda^2\tilde\rho_0 M^d_2(k,2\tilde\lambda\tilde\rho_0;\tilde\eps)}{1-36\tilde\lambda^2\tilde\rho_0 M^d_1(k,2\tilde\lambda\tilde\rho_0;\tilde\eps)} ,
\label{floweq_1}
\eleq
where the threshold functions $L^d_n$ and $M^d_{1,2}$ are defined in Appendix \ref{app_threshold}, and 
\bleq
\dt \tilde\eps(\q) &=& (\eta-2) \tilde\eps(\q) \nonumber \\ &&  -9\tilde\lambda^2\tilde\rho_0 k^{-d} \int_{\q'}  \Bigl\lbrace \left[(\eta r+2\tilde\eps_0r')\tilde\eps_0g^2\right]_{\q'} g(\q'+\q) \nonumber \\ &&
+g(\q') \left[(\eta r+2\tilde\eps_0r')\tilde\eps_0g^2\right]_{\q'+\q} \nonumber \\ && 
-2 \left[(\eta r+2\tilde\eps_0r')\tilde\eps_0g^2\right]_{\q'} g(\q') \Bigr\rbrace ,
\label{floweq_2}
\eleq 
where $r\equiv r(\tilde\eps_0(\q))$ and $r'\equiv r'(\tilde\eps_0(\q))$. We have introduced the running anomalous dimension 
\beq
\eta = - \dt \ln Z
\label{eta_def} 
\eeq
and the dimensionless propagator
\beq
g(\q) = \frac{1}{\tilde \eps(\q) + \tilde\eps_0(\q) r(\tilde\eps_0(\q))+2\tilde\lambda \tilde\rho_0} .
\label{gq_def}
\eeq
A direct consequence of the property (\ref{Rlimit}) is that the flow equations (\ref{floweq_1}) become identical to those of the continuum model in the limit $k\ll 1$, 
\bleq
\dt \tilde\rho_0 &=& -(d-2+\eta)\tilde\rho_0 +6v_d l^d_1(2\tilde\lambda\tilde\rho_0,\eta) , \nonumber \\ 
\dt \tilde\lambda &=& (d-4+2\eta)\tilde\lambda + 18 v_d \tilde\lambda^2 l^d_2(2\tilde\lambda\tilde\rho_0,\eta) , \nonumber \\ 
\eta &=& 72\frac{v_d}{d} \tilde\lambda^2\tilde\rho_0 m^d(2\tilde\lambda\tilde\rho_0,\eta) , 
\eleq
where $l^d_n$ and $m^d$ are the usual threshold functions (Appendix \ref{app_threshold}),
whereas (\ref{floweq_2}) becomes identical to the self-energy equation derived in Ref.~\onlinecite{Sinner08}. The coefficient $v_d$ is defined in Appendix \ref{app_threshold}. 

\subsection{Approximations}

Even if one takes advantage of the symmetry properties of the dispersion $\eps(\q)$ (\eg invariance under $q_\nu\to -q_\nu$ or $q_\nu \leftrightarrow q_{\nu'}$), the numerical solution of the flow equations will be quite demanding in particular for $d\geq 3$. In this section, we discuss two approximations which make the numerical solution much easier. Their reliability will be discussed in Sec.~\ref{sec_results}. 

\subsubsection{LPA'} 

The numerical solution of the flow equations (\ref{floweq_1}) and (\ref{floweq_2}) gives a complete description of the behavior of the $\phi^4$ theory (\ref{action}) in the approximation where the effective action $\Gamma[M]$ is given by (\ref{eff_action}). It yields not only the critical exponents but also the propagator $G(\q)=\mean{\phi_\q\phi_{-\q}}$ over the entire Brillouin zone. We know from previous works\cite{Blaizot06,Blaizot07,Sinner08} that the anomalous dimension $\eta$ can be obtained either from $Z$ [Eq.~(\ref{eta_def})] or from the momentum dependence of the propagator $G(\q)\sim 1/|\q|^{2-\eta}$ in the limit $\q\to 0$. In order to simplify the numerical solution, we can give up the exact solution of the flow equation (\ref{floweq_2}) satisfied by $\tilde\eps(\q)$, in particular near $\q=0$, since we still have the possibility to extract $\eta$ from the scale dependence of the wave-function renormalization factor $Z$ [Eq.~(\ref{eta_def})]. 

In the simplest approximation, one sets $\tilde\eps(\q)=\tilde\eps_0(\q)$, \ie $\eps(\q)=Z\eps_0(\q)$. For $\eps_k\ll \eps_0$, this yields $\eps(\q)=Z\eps_0\q^2$, which is nothing but the LPA' previously introduced for continuum models. The LPA' amounts to writing the effective action $\Gamma[M]$ in terms of a local potential $U$ (with a possible expansion in field truncated to a finite order) and a gradient term $Z(\nablabf M)^2$ whose amplitude is given by $Z$. As it is based on a gradient expansion, the LPA' is valid at small momentum $|\q|\lesssim k$ and gives only the long-distance behavior of the propagator. It is made possible by the fact that only propagators with momenta $|\q|\lesssim k$ enter the flow equations. In the lattice case we are considering here, the approximation $\tilde\eps(\q)=\tilde\eps_0(\q)$ can be seen as the natural extension of the LPA' used in continuum models. In the limit $\eps_k\ll\eps_0$ where the lattice plays no role any more, its accuracy will be identical to that of the LPA' in continuum models. When $\eps_k\gg \eps_0$, fluctuations are local in space so that the dispersion should not be significantly renormalized: $Z\simeq 1$ and $\eps(\q)\simeq \eps_0(\q)$. Whether or not the LPA' is also reliable when $\eps_k \sim \eps_0$ is more difficult to assess without comparing to a more complete solution of the flow equations. This will be done in Sec.~\ref{sec_results}. The flow equations in the LPA' reduce to (\ref{floweq_1}) where the threshold functions $L^d_n$ and $M^d_{1,2}$ should be computed with the replacement $\tilde\eps(\q)\to\tilde\eps_0(\q)$. 

\subsubsection{Circular harmonic expansion} 
\label{subsubsec:harmonic}

It is possible to go beyond the LPA' and obtain the renormalized dispersion over the entire Brillouin zone by expanding $\eps(\q)$ in circular harmonics,
\beq
\eps(\q) = \sum_{n,m,l=0}^L \eps_{nml} [\cos(nq_x) \cos(mq_y) \cos(lq_z) -1] ,
\label{eps_harm}
\eeq
and retaining only a subset of harmonics -- defined in (\ref{eps_harm}) by the integer $L$. In this section, we take $d=3$. For a system with nearest-neighbor interactions only [Eq.~(\ref{eps_0_def})], the initial conditions for the coefficients $\eps_{nml}$ are given by 
\bleq
\eps_{nml}\bigl|_{t=0} &=& -2\eps_0 (\delta_{n,1}\delta_{m,0}\delta_{l,0} \nonumber \\ && + \delta_{n,0}\delta_{m,1}\delta_{l,0} + \delta_{n,0}\delta_{m,0}\delta_{l,1} ) .
\eleq
The flow equations for the dimensionless amplitudes $\tilde\eps_{nml}=\eps_{nml}/(Z\eps_k)$ read (Appendix \ref{app_floweq}) 
\begin{multline}
\dt \tilde\eps_{nml} = (\eta-2) \tilde\eps_{nml} -18c_{nml}\tilde\lambda^2\tilde\rho_0 \\ \times H^{d(1)}_{nml}(k,2\tilde\lambda\tilde\rho_0,\eta;\tilde\eps) H^{d(2)}_{nml}(2\tilde\lambda\tilde\rho_0;\tilde\eps),
\label{floweq_3}
\end{multline} 
where the functions $H^{d(1)}$ and $H^{d(2)}$ are given in Appendix \ref{app_threshold}, and $c_{nml}=(2-\delta_{n,0})(2-\delta_{m,0})(2-\delta_{l,0})$. 

When $\eps_k\gg \eps_0$, the fluctuations are local and the renormalization of the dispersion $\eps(\q)$ should be negligible. When $\eps_k\lesssim \eps_0$, non-local fluctuations renormalize the dispersion and induce harmonics which are not present in the bare dispersion $\eps_0(\q)$. We expect the harmonics $\cos(nq_\nu)$ to be generated when $n\sim k^{-1}$. On the other hand, a finite value of $\tilde\eps(\q)$ acts as a mass term in the propagator (\ref{gq_def}), so that we expect the flow of $\eps(\q)$ to stop when $\tilde\eps(\q)\sim 1$, \ie when $k\sim |\q|$ for $k,|\q|\ll 1$. Thus the highest harmonics $\cos(nq_\nu)$ is of order $n\sim\min(k^{-1},|\q|^{-1})$. We deduce that the harmonics expansion (\ref{eps_harm}) is valid only if $L\gg \min(k^{-1},|\q|^{-1})$. 

In the limit $k\to 0$, any finite truncation ($L<\infty$) cannot properly describe the renormalized dispersion for $|\q|\ll L^{-1}$ and therefore the $\q\to 0$ limit of the propagator. In particular, it will always give a dispersion that behaves as $\q^2$ for $\q\to 0$ and a critical propagator $G(\q)\sim 1/\q^2$ with vanishing anomalous dimension. As in the LPA', one should then extract the anomalous dimension $\eta$ from $Z$. 

Finally, we note that we can combine the harmonic expansion with the LPA' by replacing $\tilde\eps(\q)$ by $\tilde\eps_0(\q)$ in the threshold functions $L^d_n$, $M^d_{1,2}$ and $H^{d(1,2)}$. This approximation will be referred to as the H-LPA'.

\section{Results and discussion}
\label{sec_results} 

In this section, we consider a three-dimensional system with nearest-neighbor interactions [Eq.~(\ref{eps_0_def})] and take $\eps_0=1$. 

\begin{figure}[t]
\centerline{\includegraphics[width=7cm,clip]{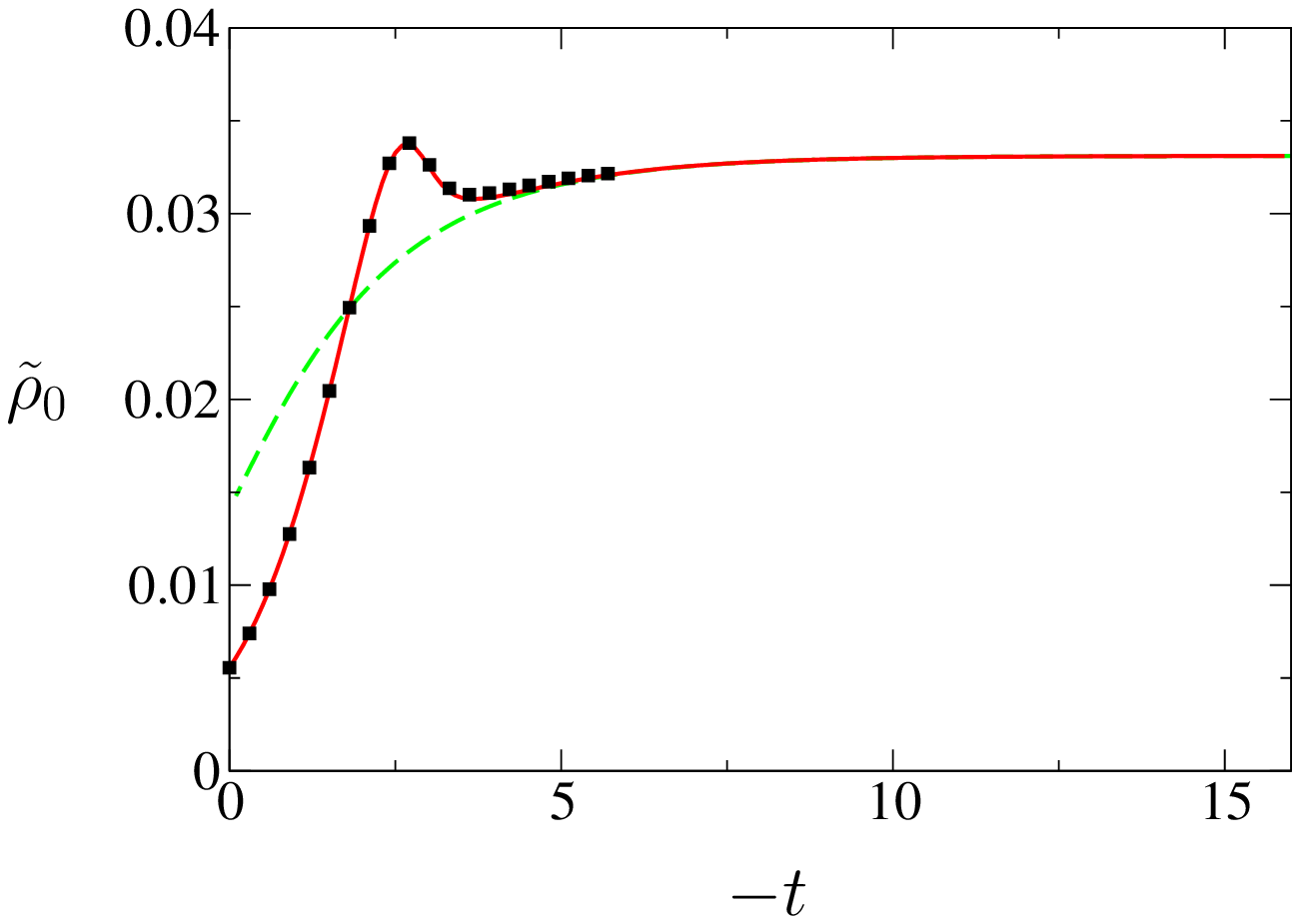}}
\caption{(Color online) $\tilde\rho_0$ {\it vs} $-t$ in the LPA' ((red) solid line) near criticality for $\lambda(t=0)=100$ ($d=3$). The black squares are obtained from the full solution of the flow equations  (\ref{floweq_1},\ref{floweq_2}). The (green) dashed line is obtained from the flow equations of the continuum model with the same boundary conditions at $t=-16$.}
\label{fig_rho_0_lpap} 
\vspace{0.25cm}
\centerline{\includegraphics[width=7cm,clip]{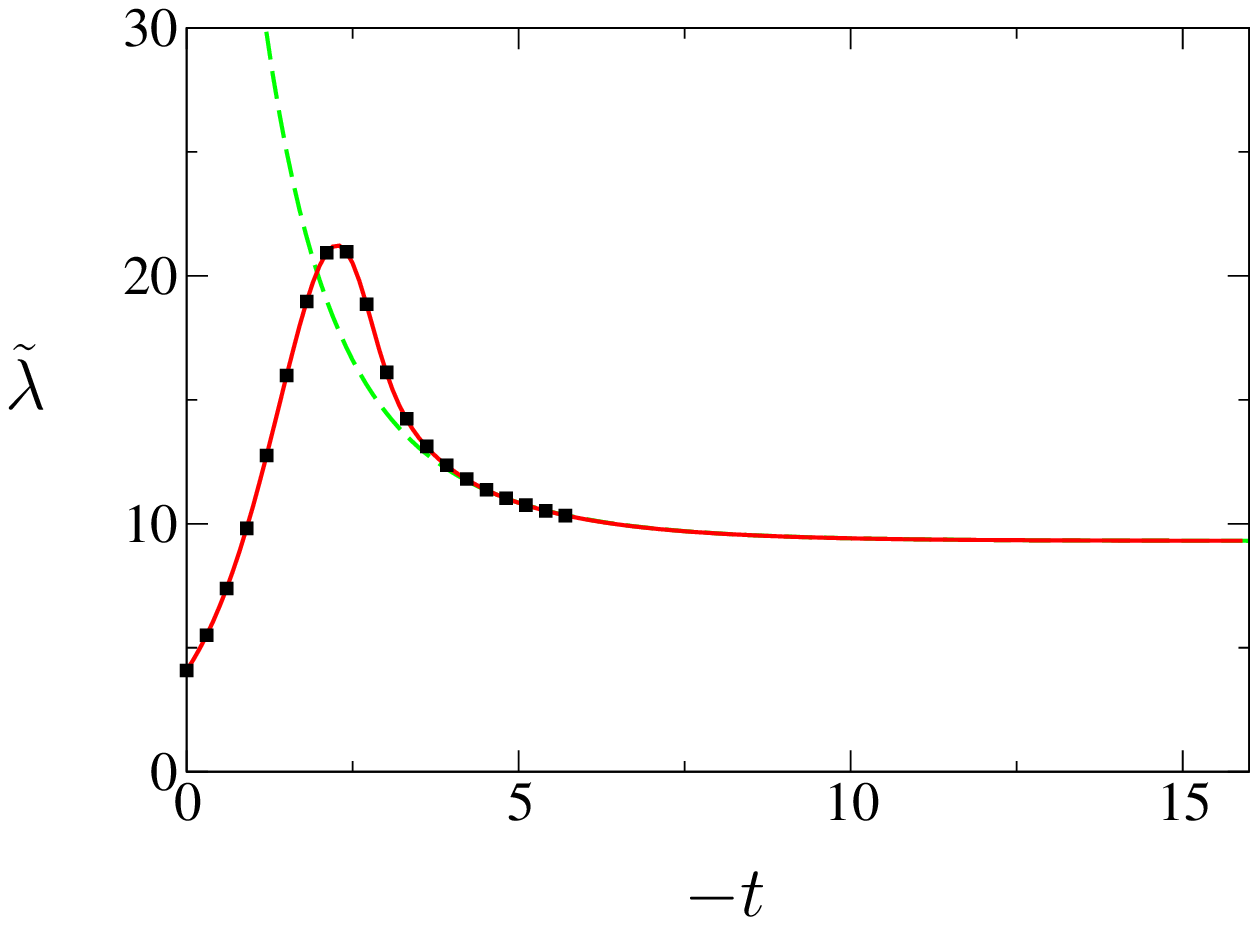}}
\caption{(Color online) Same as Fig.~\ref{fig_rho_0_lpap} but for $\lambda$. }
\label{fig_lamb_lpap} 
\vspace{0.25cm}
\centerline{\includegraphics[width=7cm,clip]{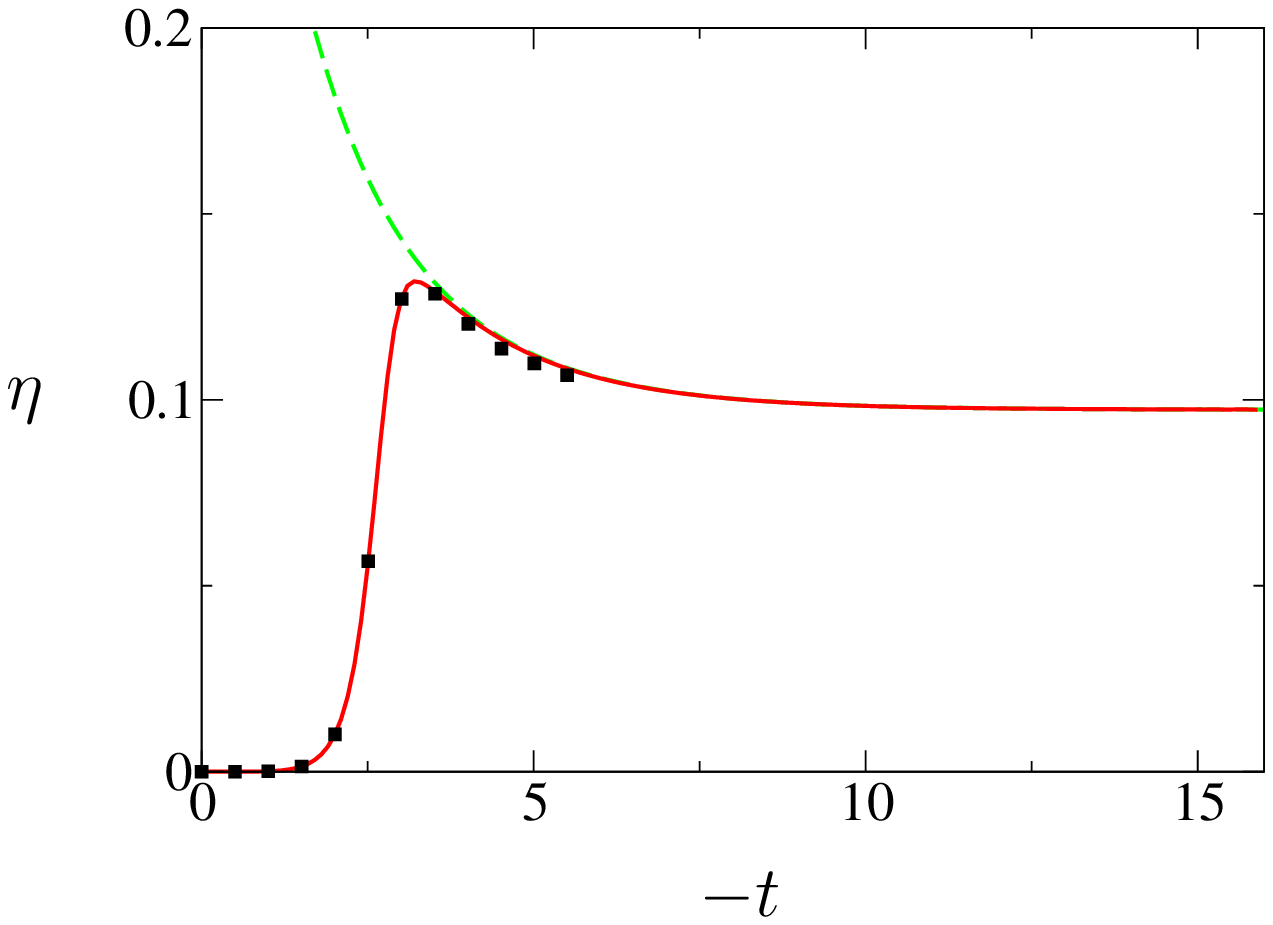}}
\caption{(Color online) Same as Fig.~\ref{fig_rho_0_lpap} but for $\eta$. }
\label{fig_eta_lpap} 
\end{figure}

Let us first discuss the case of a strong initial value $\lambda(t=0)=100$ of the interaction. The flow of $\tilde\rho_0$, $\tilde\lambda$ and $\eta$ is shown in figures \ref{fig_rho_0_lpap}, \ref{fig_lamb_lpap} and \ref{fig_eta_lpap}. The initial value $\rho_0(t=0)\simeq 0.1358$ is adjusted (to a precision of $\sim 10^{-16}$) so that the system is (nearly) critical, as shown by the plateaus observed for $|t|\gtrsim 12$ in Figs.~\ref{fig_rho_0_lpap}-\ref{fig_eta_lpap}.\cite{note5} We find a remarkable agreement between the LPA' and the full solution of the flow equations (\ref{floweq_1},\ref{floweq_2}), which shows that the LPA' is a very good approximation for any value of $\eps_k$. Due to the numerical cost, we have only attempted to solve  (\ref{floweq_1},\ref{floweq_2}) for a limited range of $t$ when no approximation is made.
Although this seems to give a slightly better estimate of the anomalous dimension (the exact value is close to 0.036), the improvement over the derivative expansion is very weak. The main limitation of our approach wrt a more accurate calculation of $\eta$ comes from the neglecting of the $\rho$ dependence of the renormalized dispersion $\epsilon(\q)$ (in particular of the wave-function renormalization\cite{Berges00} factor $Z$ defined in (\ref{Zdef})) and the simple truncation of the potential $U(\rho)$ introduced in (\ref{Udef}). 

Figures \ref{fig_rho_0_lpap}, \ref{fig_lamb_lpap} and \ref{fig_eta_lpap} also show the flow obtained for the continuum model ($\eps_0(\q)=\q^2$) with the same boundary conditions at $t=-16$. In practice, we take the final values at $t=-16$ and solve the flow equations backwards in ``time'' (\ie from $t=-16$ to 0) with the replacement $\eps_0(\q)\to \q^2$ in the threshold functions $L^d_n$ and $M^d_{1,2}$, \ie $L^d_n\to 2v_d l^d_n$ and $M^d_1 \eta + M^d_2\to m^d$ (see Appendix \ref{app_threshold}). We can clearly distinguish between a short-distance (or high-energy) regime where the lattice effects are strong and a long-distance regime ($k\lesssim k_x \sim 1$ or $|t|\gtrsim 3.2$) where the lattice effects rapidly disappear and become undetectable in the limit $k\ll k_x$.\cite{note4}

\begin{figure}
\centerline{\includegraphics[width=6.5cm,clip]{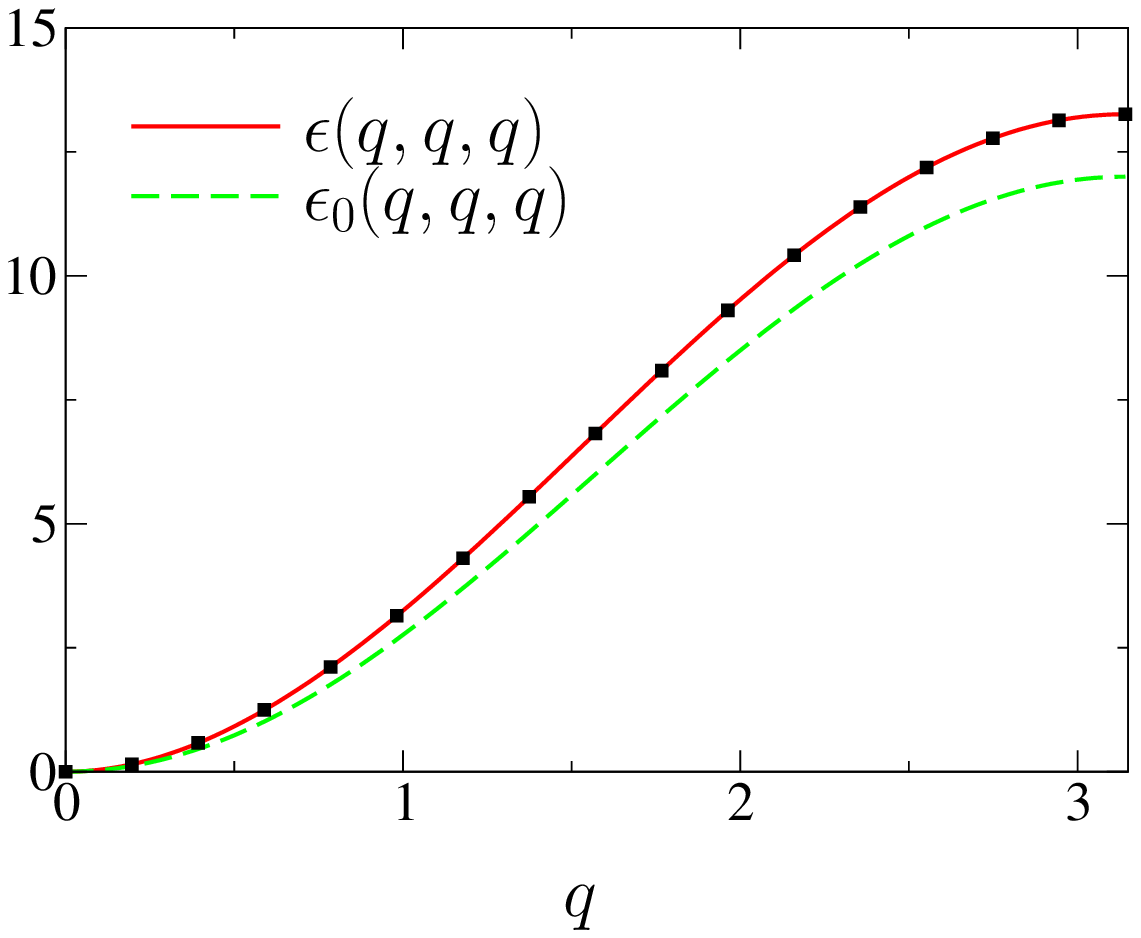}}
\caption{(Color online) Renormalized dispersion $\eps(\q)$ obtained in the H-LPA' ((red) solid line) and from the full solution of (\ref{floweq_1},\ref{floweq_2}) (black squares) along the diagonal $(0,0,0)\to (\pi,\pi,\pi)$ of the three-dimensional Brillouin zone. The bare dispersion $\eps_0(\q)$ is shown by the (green) dashed line.}
\label{fig_eps_q}
\centerline{\includegraphics[width=6.5cm,clip]{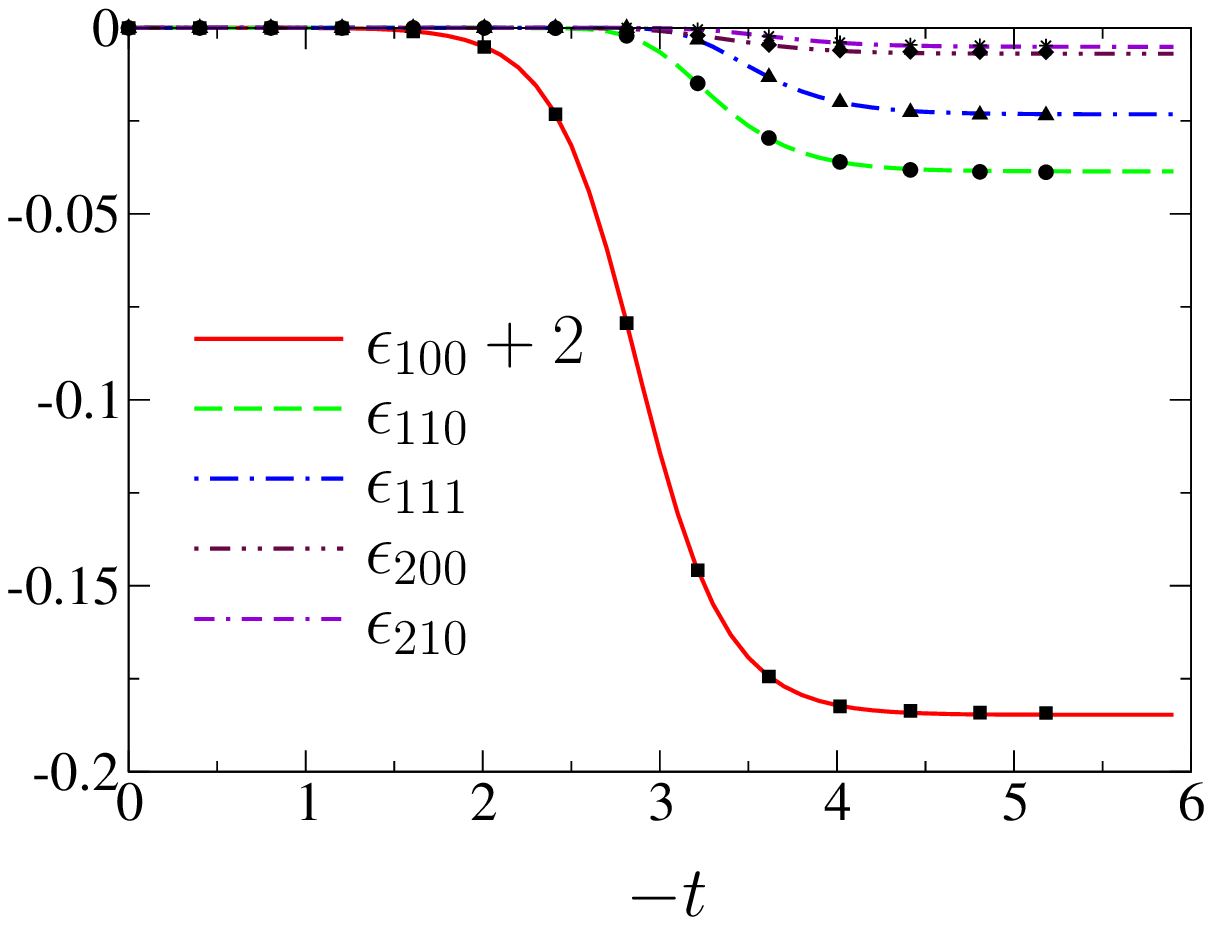}}
\caption{(Color online) Harmonic amplitudes $\eps_{nml}$ obtained in the H-LPA' (lines) and from the full solution of (\ref{floweq_1},\ref{floweq_2}) (symbols). At $t=0$, only the harmonic $\eps_{100}=\eps_{010}=\eps_{001}=-2$ is nonzero.}
\label{fig_eps_nml}
\centerline{\includegraphics[width=6.5cm,clip]{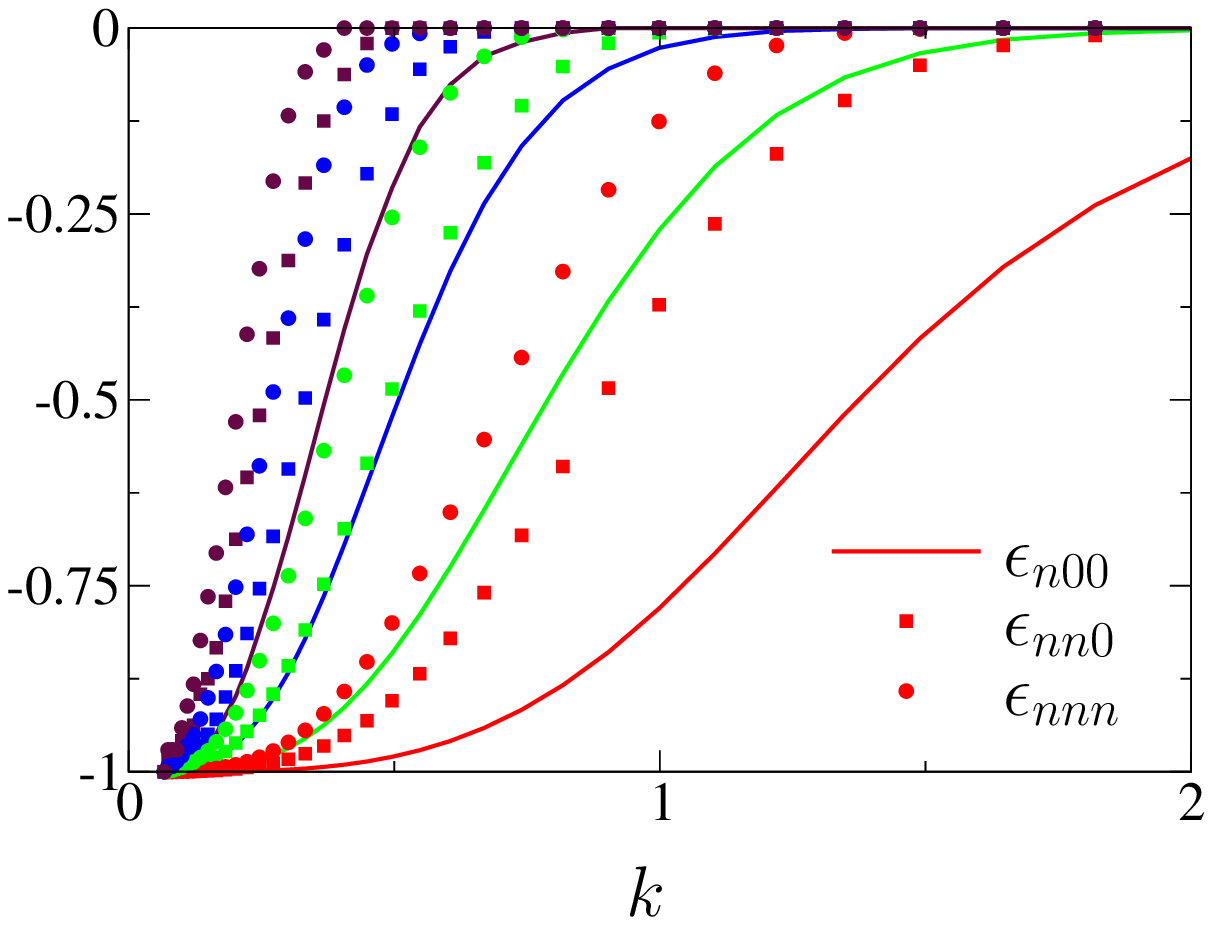}}
\caption{(Color online) Harmonic amplitudes $\eps_{n00}$ (lines), $\eps_{nn0}$ (squares) and $\eps_{nnn}$ (circles) {\it vs} $k$ for $n=1,2,3$ and 4 (from right to left). All amplitudes are normalized to their values at $k\to 0$. As in Fig.~\ref{fig_eps_nml}, $\eps_{100}$ is shifted by 2.}
\label{fig_eps_nml_k} 
\end{figure}

Given the success of the LPA' for calculating $\tilde\rho_0$, $\tilde\lambda$ and $\eta$, it makes sense to consider the H-LPA' to compute the renormalized dispersion. Figures \ref{fig_eps_q} and \ref{fig_eps_nml} show the dispersion $\eps(\q)$ as well as the first harmonics $\eps_{nml}$ ($n,m,l\leq 2$). The agreement between the H-LPA' and the full solution of the flow equations (\ref{floweq_1},\ref{floweq_2}) is again remarkable. The amplitudes $\eps_{nml}$ decrease very rapidly with $n,m,l$, and only a small number of harmonics is required for an accurate description of $\eps(\q)$ except -- as discussed in Sec.~\ref{subsubsec:harmonic} -- near $\q=0$. Note that $\rho_0$ and $\lambda$ are sensitive to local fluctuations, while $\eps(\q)$ (or $\eps_{nml}$) is not: as expected $\dt\eps_{nml}\simeq 0$ when $|t|\lesssim 2$. When $k\sim k_x$, the amplitude $\eps_{100}=\eps_{010}=\eps_{001}$ of the first harmonics varies with $k$ and higher-order harmonics are progressively generated (Figs.~\ref{fig_eps_nml} and \ref{fig_eps_nml_k}). 

\begin{figure}
\centerline{\includegraphics[width=6.5cm,clip]{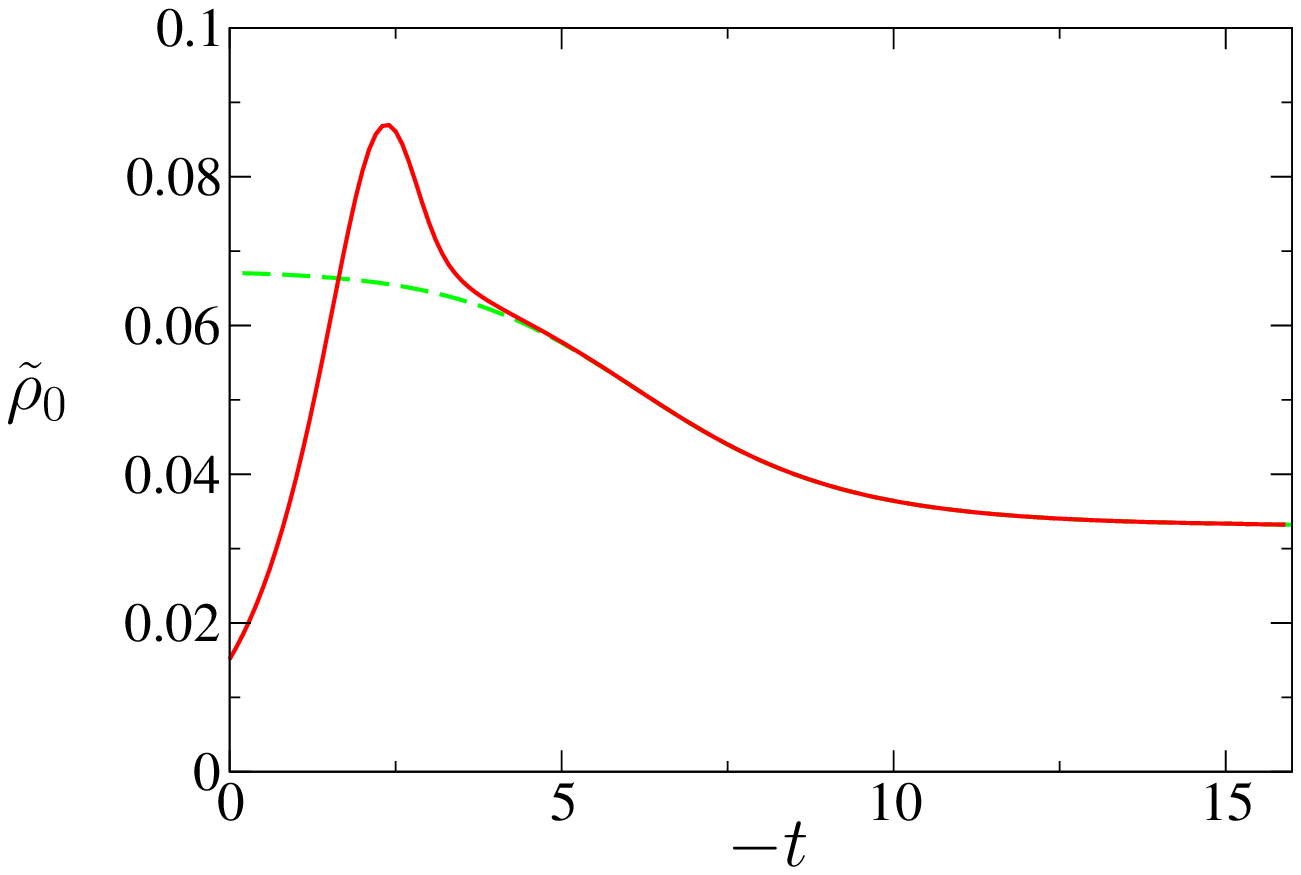}}	
\centerline{\includegraphics[width=6.5cm,clip]{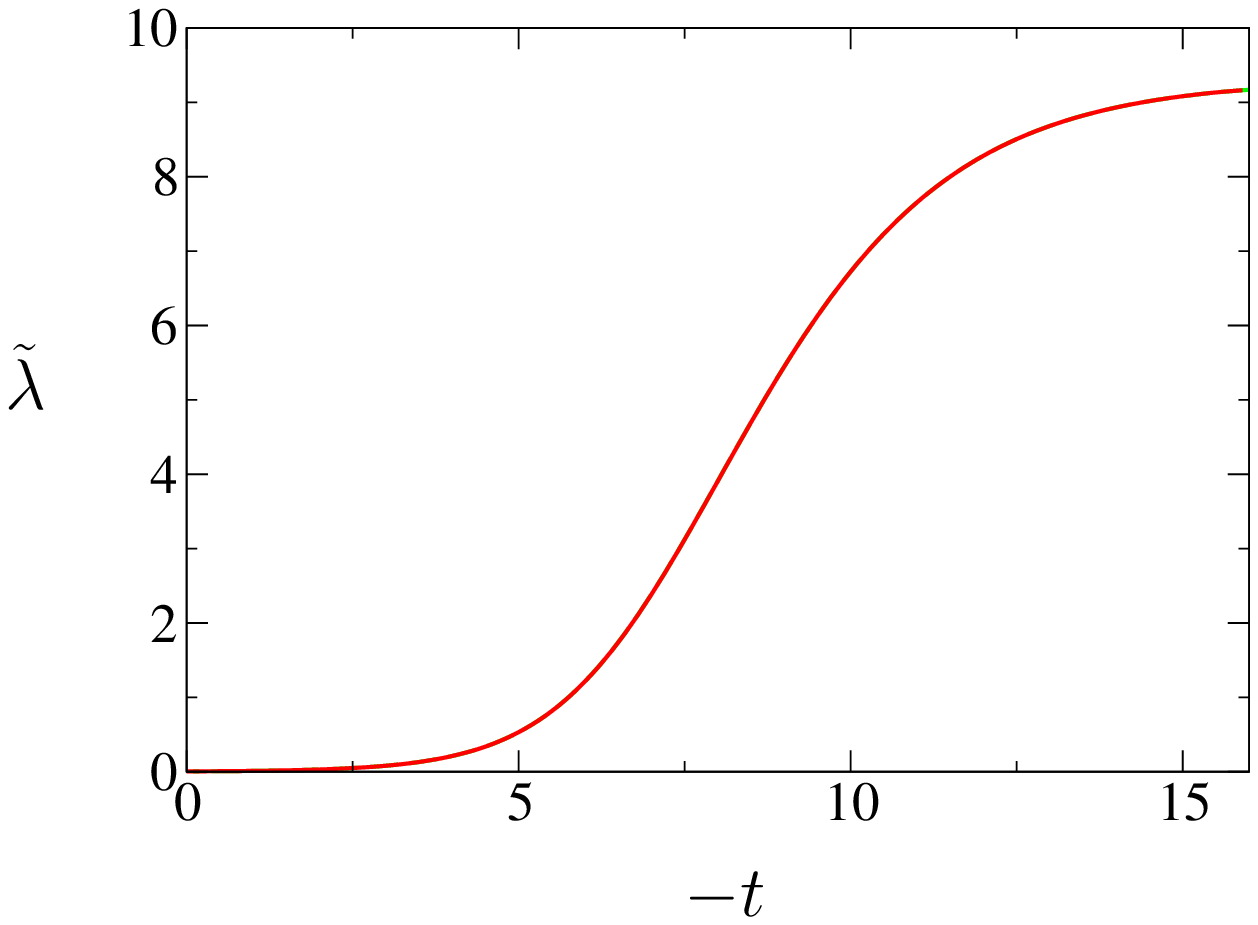}}	
\centerline{\includegraphics[width=6.5cm,clip]{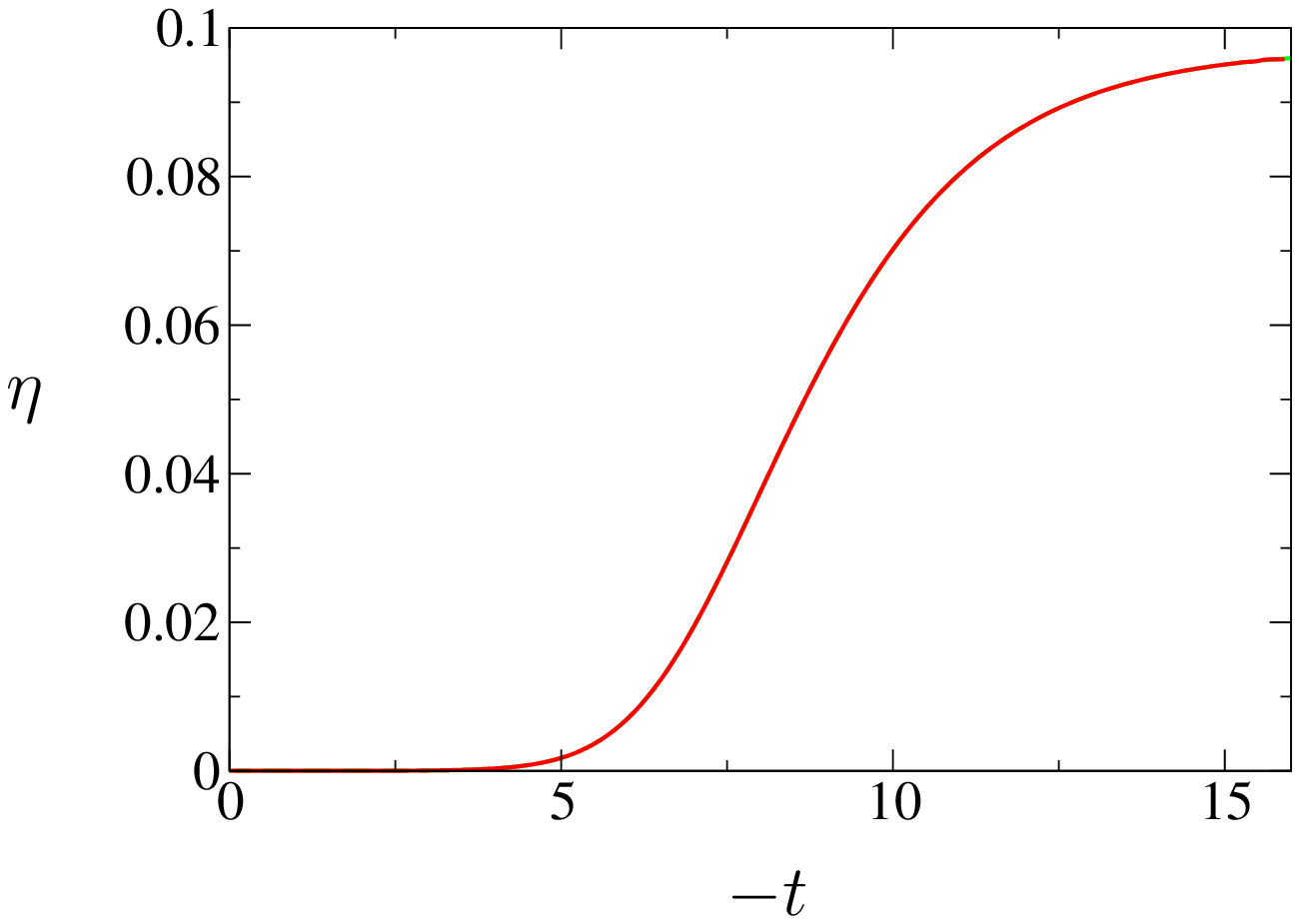}}		
\caption{Same as Figs.~\ref{fig_rho_0_lpap}, \ref{fig_lamb_lpap} and \ref{fig_eta_lpap} but for the initial condition $\lambda(t=0)=0.1$.}
\label{fig_flow_2}
\end{figure}

In Fig.~\ref{fig_flow_2}, we show the flow of $\tilde\rho_0$, $\tilde\lambda$ and $\eta$ for a weaker value of the bare coupling, $\lambda(t=0)=0.1$. To understand the dependence of the results on $\lambda(t=0)$, one should introduce the Ginzburg scale $k_c$, which is proportional to $\lambda(t=0)$ in three dimensions.\cite{Blaizot07,Hasselmann07,Sinner08} In continuum models, the Ginzburg scale separates the infrared region $\xi^{-1} \ll |\q|\ll k_c$ (with $\xi$ the correlation length) characterized by the scaling form $\Gamma^{(2)}(\q)\sim |\q|^{2-\eta^*}$ of the inverse propagator, where $\eta^*=\lim_{t\to-\infty}\eta$ is the anomalous dimension, from a perturbative regime $k_c\ll |\q| \ll \Lambda_0$. The microscopic cutoff $\Lambda_0$ should be much larger than $k_c$ for the perturbative regime to be observable. In the regime $k_c\ll k\ll \Lambda_0$, the running anomalous dimension is nearly zero and the flow is dominated by the Gaussian fixed point. Figures \ref{fig_rho_0_lpap}, \ref{fig_lamb_lpap} and \ref{fig_eta_lpap} ($\lambda(t=0)=100$) correspond to the case where $k_c\gg k_x$. As soon as the lattice scale $k_x$ is reached ($|t|\sim 3.2$), the flow rapidly crosses over to the critical regime characterized by the fixed point values $\tilde\rho_0^*$, $\tilde\lambda^*$ and the anomalous dimension $\eta^*$. Note that the finite value of $\eta(t=0)$ obtained from the continuum model (green dashed line in Fig.~\ref{fig_eta_lpap}) is explained by the fact that even at $t=0$ the system is not in the perturbative regime (\ie $k(t=0) \lesssim k_c$) for $\lambda(t=0)=100$. On the other hand, for $\lambda(t=0)=0.1$ (Fig.~\ref{fig_flow_2}), one has $k_c\ll k_x$. Between the local fluctuation regime and the critical regime, one can clearly observe an intermediate regime $k_c\ll k \ll k_x$ where the running anomalous dimension is nearly zero and the running coupling constants $\tilde\rho_0$ and $\tilde\lambda$ significantly differ from their fixed point values $\tilde\rho_0^*$ and $\tilde\lambda^*$. Note that since the renormalization of $\tilde\lambda$ and $\eta$ is nearly zero in the local fluctuation regime ($k\lesssim k_x$), the flow of these quantities is well approximated by the continuum model equations for all values of $k$ (hence the superposition of the (red) solid and (green) dashed lines in the corresponding figures). 

\section{Conclusion and perspectives}

We have shown how the presence of a lattice can be taken into account in the NPRG. Our approach allows one to compute both critical exponents and non-universal quantities such as the critical temperature or the renormalized dispersion. We have proposed two approximations which considerably reduce the numerical difficulty of solving the flow equations. While the LPA' is sufficient to obtain the small-momentum behavior of the propagator, the H-LPA' -- which is based on a circular harmonic expansion of the dispersion -- yields the renormalized dispersion over the entire Brillouin zone except at very small momenta (as in the LPA', the small momentum behavior of the propagator is deduced form the scale dependence of the wave-function renormalization factor $Z$). It is straightforward to generalize our approach to more complicated truncations of the effective action than the one considered in this paper [Eq.~(\ref{eff_action})] as well as to quantum-mechanical many-body systems.

\begin{acknowledgments} 
ND would like to thank B. Delamotte and D. Mouhanna for discussions and advice regarding the numerical solution of the flow equations. 
\end{acknowledgments}

\appendix

\section{Flow equations}
\label{app_floweq}

The flow equations for the potential $U(\rho)$ and the dispersion $\eps(\q)$ can be deduced from the flow equation of the effective action $\Gamma[M]$ and the vertex $\Gamma^{(2)}$ in a uniform field $M_\r=M=\const$. Since $\eps(\q=0)=0$, one has 
\bleq
\Gamma[M]\Bigl|_{M_\r=M} &=& N U(\rho) , \nonumber \\ 
\Gamma^{(2)}(\q,\q')\Bigl|_{M_\r=M} &=& \delta_{\q+\q',0} \left[\eps(\q)+ U'+2\rho U'' \right] , \nonumber \\ && 
\eleq
where $U'=\partial_\rho U$ and $U''=\partial^2_\rho U$. From (\ref{eq_exact}), one then finds
\beq
\dt U(\rho) = \half \int_\q G(\q) \dot R(\q) ,
\eeq
where 
\bleq
G(\q) &=& \left(\Gamma^{(2)}(\q,-\q)+R(\q) \right)^{-1} \nonumber \\ 
&=& \frac{1}{\eps(\q)+R(\q)+U'+2\rho U''}
\eleq
is the propagator in a uniform field. Using $\dot R(\q) = - Z\eps_0(\q) [\eta r+2\tilde\eps_0(\q) r']$ and $\eta$ defined by (\ref{eta_def}), one obtains
\bleq
\dt U &=& - \half \int_\q \tilde\eps_0(\q)[\eta r+2\tilde\eps_0(\q) r'] g(\q) \nonumber \\ 
&=& k^d L^d_0(k,\tilde U'+2\tilde\rho \tilde U'',\eta;\tilde\eps) ,
\eleq
where we have introduced the dimensionless propagator
\beq
g(\q) = Z\eps_k G(\q) = \frac{1}{\tilde\eps(\q)+\tilde\eps_0(\q) r + \tilde U' + 2\tilde\rho \tilde U''}.
\eeq
The lattice threshold function $L^d_0$ is defined in appendix \ref{app_threshold}. The dimensionless potential $\tilde U(\tilde\rho) = k^{-d}U(\rho)$ satisfies
\beq
\dt \tilde U = -d \tilde U +(d-2+\eta) \tilde\rho \tilde U' + L^d_0(k,\tilde U'+2\tilde\rho\tilde U'',\eta;\tilde\eps) . 
\label{Utilde_eq}
\eeq
With the truncation (\ref{Udef}), we obtain the flow equation of $\tilde\rho_0$ from the condition $\tilde U'(\tilde\rho_0)= 0$, \ie
\beq
\dt \tilde U'(\tilde\rho_0) = \dt \tilde U'\Bigl|_{\tilde\rho_0} + \tilde U''(\tilde\rho_0) \dt\tilde\rho_0 = 0 ,
\eeq
which gives the first of equations (\ref{floweq_1}). Similarly, the flow equation of $\tilde\lambda$ is derived from $\tilde\lambda=\tilde U''(\tilde\rho_0)$, \ie
\beq
\dt \tilde\lambda = \dt \tilde U''\Bigl|_{\tilde\rho_0}+ \tilde U^{(3)}(\tilde\rho_0) \dt \tilde \rho_0 
\eeq
(with $\tilde U^{(3)}=0$), which gives the second of equations (\ref{floweq_1}).
(The flow equations of $\tilde U'$ and $\tilde U''$ are deduced from (\ref{Utilde_eq}) using (\ref{Ln_prop}).) 

The flow equation (\ref{eq_exact}) of the effective action implies that the vertex $\Gamma^{(2)}(\q,\q')=\delta_{\q+\q',0}\Gamma^{(2)}(\q)$ satisfies
\begin{multline}
\dt \Gamma^{(2)}(\q) = \half \tilde\dt \sum_{\q'} G(\q') \Gamma^{(4)}(\q,-\q,\q',-\q') \\ 
- \half \tilde\dt \sum_{\q'} G(\q')G(\q'+\q) \Gamma^{(3)}(\q,\q',-\q'-\q) \\ \times \Gamma^{(3)}(-\q,\q'+\q,-\q') 
\label{gamma2_eq}
\end{multline}
in a uniform field $M_\r=M$. $\tilde\partial_t \equiv (\partial R/\partial t) \partial_R$ acts only on the regulator function $R$. The vertices in (\ref{gamma2_eq}) are obtained by differentiating (\ref{eff_action}) and setting $M_\r=M=\sqrt{2\rho}$,
\bleq
\Gamma^{(3)}(\q_1,\q_2,\q_3) &=& \delta_{\q_1+\q_2+\q_3,0} \frac{3\lambda \sqrt{2\rho}}{\sqrt{N}} , \nonumber \\ 
\Gamma^{(4)}(\q_1,\q_2,\q_3,\q_4) &=& \delta_{\q_1+\q_2+\q_3+\q_4,0} \frac{3\lambda}{N} .
\eleq
Since $\eps(\q)=\Gamma^{(2)}(\q)-\Gamma^{(2)}(\q=0)$, one deduces
\beq
\dt \eps(\q) = - 9\lambda^2\rho_0 \tilde\dt \int_{\q'} \left[ G(\q') G(\q'+\q)-G(\q')^2 \right] .
\eeq
In dimensionless form, we obtain
\begin{multline}
\dt \tilde\eps(\q) = (\eta-2) \tilde\eps(\q) \\ -9\tilde\lambda^2\tilde\rho_0 k^{-d} \tilde \dt \int_{\q'} \left[g(\q') g(\q'+\q) - g(\q')^2 \right] ,
\label{eps_eq} 
\end{multline}
where the propagator $g(\q)$ is evaluated at $\tilde\rho=\tilde\rho_0$.  Using 
\beq
\tilde\dt = \dot R \frac{\partial}{\partial R} = - (\eta r+ 2\tilde\eps_0 r') \frac{\partial}{\partial r} ,
\label{Rder}
\eeq
we obtain (\ref{floweq_2}). 

The equation for the anomalous dimension is obtained by expanding (\ref{eps_eq}) to order $\q^2$ using $\tilde\eps_0(\q),\tilde\eps(\q) \to \q^2/k^2$ for $\q\to 0$,
\beq
\eta = - \frac{9}{2} \tilde\lambda^2\tilde\rho_0 k^{2-d} \tilde\dt \int_\q \left[ \partial_{q_x} g(\q) \right]^2 .
\eeq
With (\ref{Rder}), we deduce 
\begin{multline}
\eta = - 9\tilde\lambda^2\tilde\rho_0 k^{2-d} \int_\q \left[ \partial_{q_x} g(\q) \right] \\ \times \partial_{q_x} \left[(\eta r+ 2\tilde\eps_0 r')\tilde\eps_0(\q) g(\q)^2 \right] ,
\end{multline}
which gives the last of equations (\ref{floweq_1}) when expressed in terms of the threshold functions $M^d_1$ and $M^d_2$ defined in appendix \ref{app_threshold}.

If one expands the dispersion in circular harmonics as in (\ref{eps_harm}), one has
\bleq
\dt \eps_{nml} &=& c_{nml} \int_\q \cos(nq_x) \cos(mq_y) \cos(lq_z) \dt \eps(\q) 
\nonumber \\ 
&=& -18 c_{nml} Z\eps_k \tilde\lambda^2\tilde\rho_0 k^{-d} g(\r_{nml}) \tilde\dt g(\r_{nml}) , \nonumber \\ &&
\eleq
where $c_{nml}$ is defined in Sec.~\ref{subsubsec:harmonic} and $\r_{nml}$ denotes the position of the lattice site with coordinates $(n,m,l)$. $g(\r)$ is given by the Fourier transform of the dimensionless propagator (\ref{gq_def}). Using (\ref{Rder}) to compute $\tilde\dt g(\r_{nml})$, one eventually obtains (\ref{floweq_3}).

\section{Lattice threshold functions}
\label{app_threshold}

\subsection{Definition}

The lattice threshold functions are defined by 
\bleq
L^d_n(k,w,\eta;\tilde\eps) &=& - (n+\delta_{n,0}) \frac{k^{-d}}{2} \nonumber \\ && \times \int_\q  \tilde\eps_0(\q)[\eta r+2\tilde\eps_0(\q)r'] g(\q)^{n+1} , \nonumber \\ 
M^d_1(k,w;\tilde\eps) &=&  -\frac{k^{2-d}}{4} \int_\q [\partial_{q_x} g(\q)] \partial_{q_x} \left[\tilde\eps_0(\q)r g(\q)^2\right]  \nonumber \\ 
M^d_2(k,w;\tilde\eps) &=&  -\frac{k^{2-d}}{2} \int_\q [\partial_{q_x} g(\q)] \partial_{q_x} \left[\tilde\eps_0^2(\q)r' g(\q)^2\right] , \nonumber \\
H^{d(1)}_{nml}(k,w,\eta;\tilde\eps) &=& k^{-d} \int_\q e^{i\q\cdot\r_{nml}} 
[\eta r+2\tilde\eps_0(\q)r'] \nonumber \\ && \times \tilde\eps_0(\q) g(\q)^2 , \nonumber \\ H^{d(2)}_{nml}(w;\tilde\eps) &=& \int_\q e^{i\q\cdot\r_{nml}} g(\q) ,
\label{thereshold_dn}
\eleq
where $g(\q)=1/(\tilde\eps(q)+\tilde\eps_0(q)r+w)$. The functions $L^d_n$ satisfy
\beq
\partial_w L^d_n(k,w,\eta;\tilde\eps) = -(n+\delta_{n,0}) L^d_{n+1}(k,w,\eta;\tilde\eps) . 
\label{Ln_prop}
\eeq

\subsection{$k\to 0$ limit} 

In the limit $k\to 0$, the functions $r$ and $r'$ ensures that the integrals determining $L^d_n$ and $M^d_{1,2}$ are dominated by $|\q|\lesssim k\ll 1$. One can then use $\tilde\eps_0(\q) \simeq y$ and $\tilde\eps(\q) \simeq y$ ($y=\q^2/k^2$) and replace the integral over the Brillouin zone by
\beq
\int_{-\infty}^\infty \frac{dq_1}{2\pi} \cdots \int_{-\infty}^\infty \frac{dq_q}{2\pi} = 
2v_d k^d \int_0^\infty dy\, y^{d/2-1} ,
\eeq
where $v_d^{-1}=2^{d+1}\pi^{d/2}\Gamma(d/2)$. This gives
\bleq
& \lim_{k\to 0} L^d_n(k,w,\eta;\tilde\eps) = 2v_d l^d_n(w,\eta) , & \nonumber \\  
& \lim_{k\to 0} \left[M^d_1(k,w;\tilde\eps) \eta + M^d_2(k,w;\tilde\eps) \right] = \dfrac{2v_d}{d} m^d(w,\eta) , & \nonumber \\ && 
\eleq
where
\bleq
l^d_n(w,\eta) &=& - \half (n+\delta_{n,0}) \int_0^\infty dy\, y^{d/2} (\eta r+2yr') g^{n+1} , \nonumber \\ 
m^d(w,\eta) &=& \int_0^\infty dy\, y^{d/2} (1+r+yr') g^4 \nonumber \\ 
&& \times \lbrace [\eta r+(\eta+4)yr'+2y^2r''] \nonumber \\ 
&& - 2yg(1+r+yr')(\eta r+2yr') \rbrace 
\eleq
($g=1/(y(1+r(y))+w)$) are the usual threshold functions of the continuum model. Note that in the $k\to 0$ limit, the threshold functions $L^d_n$ and $M^d_{1,2}$ become independent of $k$ and $\tilde\eps(\q)$.  


\begin{thebibliography}{22}
\expandafter\ifx\csname natexlab\endcsname\relax\def\natexlab#1{#1}\fi
\expandafter\ifx\csname bibnamefont\endcsname\relax
  \def\bibnamefont#1{#1}\fi
\expandafter\ifx\csname bibfnamefont\endcsname\relax
  \def\bibfnamefont#1{#1}\fi
\expandafter\ifx\csname citenamefont\endcsname\relax
  \def\citenamefont#1{#1}\fi
\expandafter\ifx\csname url\endcsname\relax
  \def\url#1{\texttt{#1}}\fi
\expandafter\ifx\csname urlprefix\endcsname\relax\def\urlprefix{URL }\fi
\providecommand{\bibinfo}[2]{#2}
\providecommand{\eprint}[2][]{\url{#2}}

\bibitem[{\citenamefont{Wilson and Kogut}(1974)}]{Wilson74}
\bibinfo{author}{\bibfnamefont{K.~G.} \bibnamefont{Wilson}} \bibnamefont{and}
  \bibinfo{author}{\bibfnamefont{J.~B.} \bibnamefont{Kogut}},
  \bibinfo{journal}{Phys. Rep.} \textbf{\bibinfo{volume}{12}},
  \bibinfo{pages}{75} (\bibinfo{year}{1974}).

\bibitem[{\citenamefont{Polchinski}(1984)}]{Polchinski84}
\bibinfo{author}{\bibfnamefont{J.}~\bibnamefont{Polchinski}},
  \bibinfo{journal}{Nucl. Phys. B} p. \bibinfo{pages}{231}
  (\bibinfo{year}{1984}).

\bibitem[{\citenamefont{Wetterich}(1993)}]{Wetterich93}
\bibinfo{author}{\bibfnamefont{C.}~\bibnamefont{Wetterich}},
  \bibinfo{journal}{Phys. Lett. B} \textbf{\bibinfo{volume}{301}},
  \bibinfo{pages}{90} (\bibinfo{year}{1993}).

\bibitem[{\citenamefont{Berges et~al.}(2000)\citenamefont{Berges, Tetradis, and
  Wetterich}}]{Berges00}
\bibinfo{author}{\bibfnamefont{J.}~\bibnamefont{Berges}},
  \bibinfo{author}{\bibfnamefont{N.}~\bibnamefont{Tetradis}}, \bibnamefont{and}
  \bibinfo{author}{\bibfnamefont{C.}~\bibnamefont{Wetterich}},
  \bibinfo{journal}{Phys. Rep.} \textbf{\bibinfo{volume}{363}},
  \bibinfo{pages}{223} (\bibinfo{year}{2000}).

\bibitem[{\citenamefont{Delamotte}()}]{Delamotte07}
\bibinfo{author}{\bibfnamefont{B.}~\bibnamefont{Delamotte}},
  \bibinfo{note}{arXiv:cond-mat/0702365}.

\bibitem[{not({\natexlab{a}})}]{note1}
\bibinfo{note}{The derivative expansion is a good approximation to the vertices
  only when the external momenta are smaller than the lowest mass in the
  problem. For massless theories, it therefore provides information only about
  the vertices at vanishing momenta.}

\bibitem[{\citenamefont{Blaizot
  et~al.}(2006{\natexlab{a}})\citenamefont{Blaizot, M\'endez-Galain, and
  Wschebor}}]{Blaizot06}
\bibinfo{author}{\bibfnamefont{J.-P.} \bibnamefont{Blaizot}},
  \bibinfo{author}{\bibfnamefont{R.}~\bibnamefont{M\'endez-Galain}},
  \bibnamefont{and} \bibinfo{author}{\bibfnamefont{N.}~\bibnamefont{Wschebor}},
  \bibinfo{journal}{Phys. Lett. B} \textbf{\bibinfo{volume}{632}},
  \bibinfo{pages}{571} (\bibinfo{year}{2006}{\natexlab{a}}).

\bibitem[{\citenamefont{Blaizot et~al.}(2007)\citenamefont{Blaizot,
  M\'endez-Galain, and Wschebor}}]{Blaizot07}
\bibinfo{author}{\bibfnamefont{J.-P.} \bibnamefont{Blaizot}},
  \bibinfo{author}{\bibfnamefont{R.}~\bibnamefont{M\'endez-Galain}},
  \bibnamefont{and} \bibinfo{author}{\bibfnamefont{N.}~\bibnamefont{Wschebor}},
  \bibinfo{journal}{Eur. Phys. J. B} \textbf{\bibinfo{volume}{58}},
  \bibinfo{pages}{297} (\bibinfo{year}{2007}).

\bibitem[{\citenamefont{Blaizot
  et~al.}(2006{\natexlab{b}})\citenamefont{Blaizot, M\'endez-Galain, and
  Wschebor}}]{Blaizot06a}
\bibinfo{author}{\bibfnamefont{J.-P.} \bibnamefont{Blaizot}},
  \bibinfo{author}{\bibfnamefont{R.}~\bibnamefont{M\'endez-Galain}},
  \bibnamefont{and} \bibinfo{author}{\bibfnamefont{N.}~\bibnamefont{Wschebor}},
  \bibinfo{journal}{Phys. Rev. E} \textbf{\bibinfo{volume}{74}},
  \bibinfo{pages}{051116} (\bibinfo{year}{2006}{\natexlab{b}}).

\bibitem[{\citenamefont{Blaizot
  et~al.}(2006{\natexlab{c}})\citenamefont{Blaizot, M\'endez-Galain, and
  Wschebor}}]{Blaizot06b}
\bibinfo{author}{\bibfnamefont{J.-P.} \bibnamefont{Blaizot}},
  \bibinfo{author}{\bibfnamefont{R.}~\bibnamefont{M\'endez-Galain}},
  \bibnamefont{and} \bibinfo{author}{\bibfnamefont{N.}~\bibnamefont{Wschebor}},
  \bibinfo{journal}{Phys. Rev. E} \textbf{\bibinfo{volume}{74}},
  \bibinfo{pages}{051117} (\bibinfo{year}{2006}{\natexlab{c}}).

\bibitem[{\citenamefont{Guerra et~al.}(2007)\citenamefont{Guerra,
  M\'endez-Galain, and Wschebor}}]{Guerra07}
\bibinfo{author}{\bibfnamefont{D.}~\bibnamefont{Guerra}},
  \bibinfo{author}{\bibfnamefont{R.}~\bibnamefont{M\'endez-Galain}},
  \bibnamefont{and} \bibinfo{author}{\bibfnamefont{N.}~\bibnamefont{Wschebor}},
  \bibinfo{journal}{Eur. Phys. J. B} \textbf{\bibinfo{volume}{59}},
  \bibinfo{pages}{357} (\bibinfo{year}{2007}).

\bibitem[{\citenamefont{Benitez et~al.}(2008)\citenamefont{Benitez,
  M\'{e}ndez-Galain, and Wschebor}}]{Benitez08}
\bibinfo{author}{\bibfnamefont{F.}~\bibnamefont{Benitez}},
  \bibinfo{author}{\bibfnamefont{R.}~\bibnamefont{M\'{e}ndez-Galain}},
  \bibnamefont{and} \bibinfo{author}{\bibfnamefont{N.}~\bibnamefont{Wschebor}},
  \bibinfo{journal}{Phys. Rev. B} \textbf{\bibinfo{volume}{77}},
  \bibinfo{pages}{024431} (\bibinfo{year}{2008}).

\bibitem[{\citenamefont{Sinner et~al.}(2008)\citenamefont{Sinner, Hasselmann,
  and Kopietz}}]{Sinner08}
\bibinfo{author}{\bibfnamefont{A.}~\bibnamefont{Sinner}},
  \bibinfo{author}{\bibfnamefont{N.}~\bibnamefont{Hasselmann}},
  \bibnamefont{and} \bibinfo{author}{\bibfnamefont{P.}~\bibnamefont{Kopietz}},
  \bibinfo{journal}{J. Phys.: Cond. Matt.} \textbf{\bibinfo{volume}{20}},
  \bibinfo{pages}{075208} (\bibinfo{year}{2008}).

\bibitem[{\citenamefont{Hasselmann et~al.}(2007)\citenamefont{Hasselmann,
  Sinner, and Kopietz}}]{Hasselmann07}
\bibinfo{author}{\bibfnamefont{N.}~\bibnamefont{Hasselmann}},
  \bibinfo{author}{\bibfnamefont{A.}~\bibnamefont{Sinner}}, \bibnamefont{and}
  \bibinfo{author}{\bibfnamefont{P.}~\bibnamefont{Kopietz}},
  \bibinfo{journal}{Phys. Rev. E} \textbf{\bibinfo{volume}{76}}
  (\bibinfo{year}{2007}).

\bibitem[{not({\natexlab{b}})}]{note2}
\bibinfo{note}{An example is given by the Mott-superfluid transition of
  interacting lattice bosons at commensurate density (integer number of bosons
  per site). In the absence of a lattice, the ground state is always superfluid
  regardless of the strength of the interactions between bosons. NPRG studies
  of interacting bosons in the continuum have recently been reported in C.
  Wetterich, Phys. Rev. B {\bf 77}, 064504 (2008); N. Dupuis and K. Sengupta,
  Europhys. Lett. {\bf 80}, 50007 (2007).}

\bibitem[{\citenamefont{Baier et~al.}(2004)\citenamefont{Baier, Bick, and
  Wetterich}}]{Baier04}
\bibinfo{author}{\bibfnamefont{T.}~\bibnamefont{Baier}},
  \bibinfo{author}{\bibfnamefont{E.}~\bibnamefont{Bick}}, \bibnamefont{and}
  \bibinfo{author}{\bibfnamefont{C.}~\bibnamefont{Wetterich}},
  \bibinfo{journal}{Phys. Rev. B} \textbf{\bibinfo{volume}{70}},
  \bibinfo{pages}{125111} (\bibinfo{year}{2004}).

\bibitem[{\citenamefont{Baier et~al.}(2005)\citenamefont{Baier, Bick, and
  Wetterich}}]{Baier05}
\bibinfo{author}{\bibfnamefont{B.}~\bibnamefont{Baier}},
  \bibinfo{author}{\bibfnamefont{E.}~\bibnamefont{Bick}}, \bibnamefont{and}
  \bibinfo{author}{\bibfnamefont{C.}~\bibnamefont{Wetterich}},
  \bibinfo{journal}{Phys. Lett. B} \textbf{\bibinfo{volume}{605}},
  \bibinfo{pages}{144} (\bibinfo{year}{2005}).

\bibitem[{\citenamefont{Krahl and Wetterich}(207)}]{Krahl07}
\bibinfo{author}{\bibfnamefont{H.~C.} \bibnamefont{Krahl}} \bibnamefont{and}
  \bibinfo{author}{\bibfnamefont{C.}~\bibnamefont{Wetterich}},
  \bibinfo{journal}{Phys. Lett. A} \textbf{\bibinfo{volume}{367}},
  \bibinfo{pages}{263} (\bibinfo{year}{207}).

\bibitem[{\citenamefont{Krahl et~al.}()\citenamefont{Krahl, M\"uller, and
  Wetterich}}]{Krahl08}
\bibinfo{author}{\bibfnamefont{H.~C.} \bibnamefont{Krahl}},
  \bibinfo{author}{\bibfnamefont{J.~A.} \bibnamefont{M\"uller}},
  \bibnamefont{and}
  \bibinfo{author}{\bibfnamefont{C.}~\bibnamefont{Wetterich}},
  \bibinfo{note}{arXiv:0801.1773}.

\bibitem[{not({\natexlab{c}})}]{note3}
\bibinfo{note}{See e.g. Appendix A in Ref.~\onlinecite{Blaizot07}.}

\bibitem[{not({\natexlab{d}})}]{note5}
\bibinfo{note}{The value of $\rho_0(t=0)=-3v/u=-v/\lambda(t=0)$ for which the
  system is critical determines the critical temperature $T_c$ if the
  dependence of $v$ on $T$ is known. The critical temperature can then be
  related to the parameters of the lattice model without any dependence on an
  effective short-distance cutoff. This should be contrasted to the continnum
  version of the same model where the relation between the bare parameters of
  the action (defined at an effective short-distance cutoff $\Lambda_0$) and
  the parameters of the underlying lattice model is usually unknown.}

\bibitem[{not({\natexlab{e}})}]{note4}
\bibinfo{note}{The estimates of $\tilde\rho_0(t=0)$ and $\tilde\lambda(t=0)$
  based on the continuum model flow equations run backwards in "time" from
  $t=-16$ (green lines in Figs.~\ref{fig_lamb_lpap} and \ref{fig_eta_lpap}) are
  given by 75 and 0.38, respectively.}

\end{thebibliography}

\end{document}